\documentclass{article}
\usepackage[table]{xcolor}

\usepackage{graphicx} 
\usepackage{amsmath}  
\usepackage{amsfonts} 

\usepackage[parfill]{parskip}

\usepackage[scale=0.8,vmarginratio={1:2},heightrounded]{geometry}

\usepackage{amsthm} 
\usepackage[affil-it]{authblk} 
\usepackage{amsopn} 
\usepackage{amssymb} 
\usepackage{mathrsfs} 
\usepackage{booktabs} 
\usepackage{natbib} 
\usepackage{tikz} 
\usepackage{algorithm} 
\usepackage{float} 
\usepackage{caption} 
\usepackage{abstract} 
\usepackage{listings} 

\usepackage[]{hyperref} 

\newfloat{algorithm}{t}{lop}

\numberwithin{equation}{section}

\theoremstyle{definition}

\theoremstyle{remark}

\theoremstyle{theorem}

\newcommand{\ZZ}{\mathbb{Z}}
\newcommand{\RR}{\mathbb{R}}

\newcommand{\EE}[1]{\mathbb{E} \left[ #1 \right]}

\newcommand{\de}{\mathrm{d}}

\newcommand{\var}[1]{\, {\rm var}\left( #1 \right) }

\DeclareMathOperator*{\argmin}{argmin} 

\newcommand{\LSmoments}[2]{\left|#1_#2(\omega;\theta_\omega)-\hat{#1}_#2(\omega)\right|^2}
\newcommand{\LSregress}[1]{\big(#1(\omega;\theta)-\hat#1(\omega)\big)^2}
\newcommand{\tr}{\text{tr}}

\usepackage{siunitx}
\usepackage{svg, subcaption}
\usepackage[]{hyperref}
\usepackage{mathpazo}
\usepackage{multirow}
\usepackage{arydshln}

\title{A multivariate pseudo-likelihood approach to estimating directional ocean wave models}

\author[1]{Jake P. Grainger}
\author[2]{Adam M. Sykulski}
\author[4]{Kevin Ewans}
\author[5]{Hans F. Hansen}
\author[2,3]{Philip Jonathan}
\affil[1]{STOR-i Centre for Doctoral Training, Department of Mathematics and Statistics, Lancaster University, Lancaster, UK}
\affil[2]{Department of Mathematics and Statistics, Lancaster University, Lancaster,UK} 
\affil[3]{Shell Research Ltd., London, UK} 
\affil[4]{MetOcean Research Ltd., New Plymouth, New Zealand}
\affil[5]{HAW MetOcean ApS, DK-1860 Frederiksberg C, Denmark}

\date{}

\begin{document}

\maketitle

\begin{abstract}

Ocean buoy data in the form of high frequency multivariate time series are routinely recorded at many locations in the world's oceans.
Such data can be used to characterise the ocean wavefield, which is important for numerous socio-economic and scientific reasons.
This characterisation is typically achieved by modelling the frequency-direction spectrum, which decomposes spatiotemporal variability by both frequency and direction.
State-of-the-art methods for estimating the parameters of such models do not make use of the full spatiotemporal content of the buoy observations due to unnecessary assumptions and smoothing steps.
We explain how the multivariate debiased Whittle likelihood can be used to jointly estimate all parameters of such frequency-direction spectra directly from the recorded time series. 
When applied to North Sea buoy data, debiased Whittle likelihood inference reveals smooth evolution of spectral parameters over time.
We discuss challenging practical issues including model misspecification, and provide guidelines for future application of the method.

\emph{Keywords}: Buoy displacement time-series, Debiased Whittle likelihood inference, Frequency-direction spectra, North Sea, Ocean waves, Storm spectral evolution

\end{abstract}

\section{Introduction}
Applications of multivariate time series and spatiotemporal statistics are ubiquitous, for example using the affordable and widespread availability of GPS and accelerometer technology to track individuals and objects in three spatial dimensions over time. 
Applications include clinical studies of human rest/activity cycles (actigraphy) \citep{Geraci2019}, player activity in sports \citep{Tierney2016}, motor vehicle tracking (telematics) \citep{Verbelen2018}, animal and wildlife tracking \citep{Rivest2016}, the tracking of large-scale currents and  drifting objects in oceanography \citep{Sykulski2016,OMalley2021}, as well as measuring ocean surface waves using buoys---the final of which is the focus of this paper. 
The raw time series obtained from such devices are high frequency, but often noisy, and current practices throw away or over-smooth data without utilising their full potential. 
In this paper we present a likelihood-based stochastic modelling approach that can meaningfully extract and estimate more spatiotemporal features from ocean wave observations than current methods---but we present our methodology in such a way that it can be applied more broadly to spatiotemporal data, including handling model misspecification, anisotropy, high- and low-frequency noise, aliasing, non-stationarity, and uncertainty quantification.

Wind-generated surface gravity waves are an important feature of the ocean environment.
Understanding the behaviour of such waves is of great scientific and engineering interest, with applications ranging from the design of ships and marine structures to modelling coastal flood risk.
As such, large quantities of high frequency time series data are routinely recorded in order to help improve our understanding of the waves, generating a variety of statistical challenges.
Characterisation of the wave environment needs to reflect the evolving nature of multiple weather systems, and the presence of measurement uncertainty.

From a modelling perspective, we typically seek to model the vertical displacement of the ocean surface over two-dimensional horizontal space and time.
The second-order characteristics of this spatiotemporal process are usually summarised by the frequency-direction spectrum, which ``is the fundamental quantity of wave modelling and the quantity that
allows us to calculate the consequences of interactions between waves and
other matter''\citep{Barstow2005}. 
Heuristically, the frequency-direction spectrum quantifies the contribution to the variance of the wave process from multiple sinusoidal components with different frequencies travelling in different directions.\footnote{See Appendix for a more formal definition.}
This description assumes that the wave process is stationary; however, in reality this is not generally the case.
To address this, wave records are usually partitioned into shorter intervals of time series (referred to as sea states), each of which can be treated as stationary \citep{Holthuijsen2007}.

High resolution measurements of the ocean surface in space and time are rarely available.
However, recordings of the vertical displacement of the ocean's surface at a single location (e.g. using a wave staff or downward-facing radar) or of the motions of floating devices (e.g. buoys) are common \citep{Jensen2011}. 
In particular, modern buoy measurements provide time series of the buoy's full three-dimensional displacement.
Such measurements are then used to estimate the frequency-direction spectrum, though in general this estimation is not trivial to perform.

Parametric estimation of the frequency-direction spectrum usually uses either a method-of-moments or least squares approach.
In general, neither approach is optimal statistically.
Furthermore, these techniques rely on non-parametric estimates of the frequency-direction spectrum, which exhibit substantial bias.
As a result, these approaches perform poorly and can only reliably estimate simple location parameters such as the peak frequency or mean direction of the observed wave process.
We propose using likelihood inference directly on the buoy data, avoiding both the poor performance of method-of-moments and least squares; and the issues generated by the non-parametric estimation.
Ideally, parametric inference would be made using maximum likelihood estimation with the full sample likelihood; however, the full likelihood is expensive to compute. 
Fortunately, adoption of the Whittle likelihood \citep{Whittle1953a} provides a computationally efficient alternative to full maximum likelihood inference, which produces consistent parameter estimates and is optimally convergent.
Furthermore, debiased Whittle likelihood inference \citep{Sykulski2019} removes the finite sample bias present in Whittle likelihood inference, without compromising standard error or computational efficiency.

\cite{Grainger2021} demonstrated in a univariate setting that debiased Whittle likelihood inference yields significant improvements over competitors, when estimating parameters of the spectral density function of ocean waves recorded only over time.
The paper we present here seeks to generalise this methodology to incorporate directional characteristics of the wavefield.
However, this extension is nontrivial, since the full spatiotemporal process, which constitutes the wavefield, is not recorded, and hence the spatial debiased Whittle likelihood of \cite{Guillaumin2019} cannot be applied directly.
Instead we describe computationally efficient parametric estimation of a frequency-direction spectrum fitted directly to multivariate time series buoy data.
Using a multivariate extension of the debiased Whittle likelihood we are able to obtain parameter estimates with lower bias and variance than competitor techniques. 
Our real-world data analysis reveals robust parameter estimates and captures their evolution over time during a storm; in contrast, such an analysis using existing techniques results in estimates that evolve erratically over time.

The structure of the paper is as follows.
Section~\ref{sec:background} gives some background on ocean waves, introduces an example data set and describes a model for the frequency-direction spectrum of wind-sea waves.
Section~\ref{sec:modelling} describes the debiased Whittle likelihood inference, and demonstrates its performance by simulation.
In Section~\ref{sec:example_analysis}, we then apply the debiased Whittle inference to the example data set introduced in Section~\ref{sec:background:example}, discussing a number of important practicalities of the analysis.
Finally, Section~\ref{sec:conclusion} provides a discussion and conclusions.

\newpage 
\section{Background}\label{sec:background}
\subsection{Ocean waves and frequency-direction spectra}\label{sec:background:waves}
Much of the interest in ocean waves relates to the surface displacement of the water over space and time, which is treated as a stochastic random field.
Usually, the waves are assumed to be stationary and mean-zero within a given time window (often 30 minutes), referred to as a sea state.
The covariance structure of the random field for this sea state is then described by the frequency-direction spectrum, $S(\omega,\phi)$, the frequency-domain equivalent of the spatio-temporal autocovariance (see Appendix~\ref{app:fdspec:def} for more details).

\begin{figure}[h]
    \centering
    \includegraphics[width = \linewidth]{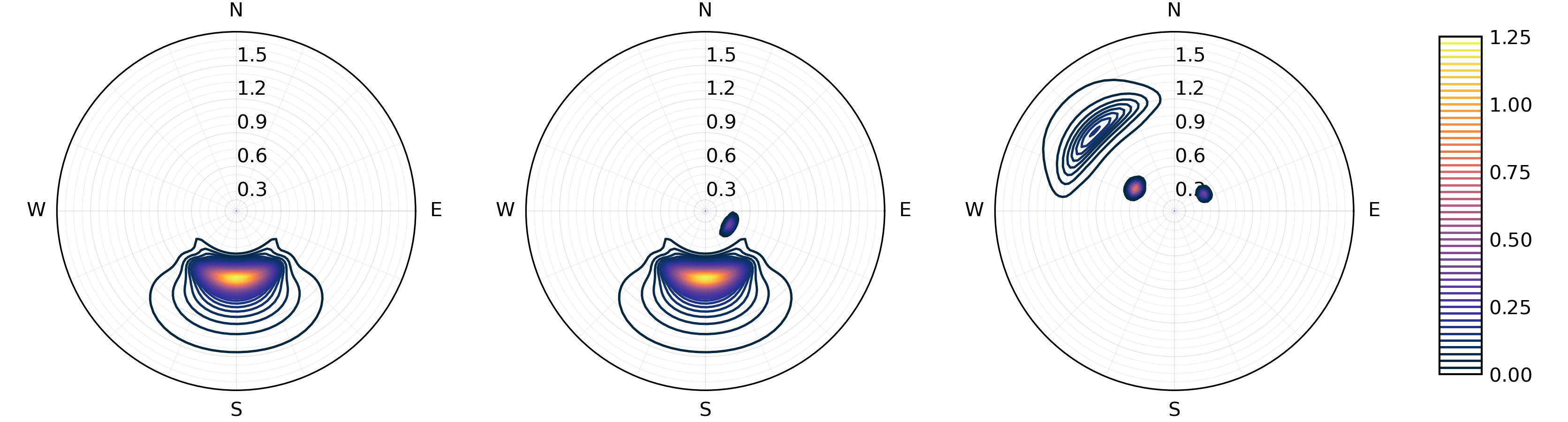}
    \caption{Example frequency-direction spectra. The left hand panel shows the frequency-direction spectra corresponding to a single wind-sea, the middle shows a wind-sea and single swell and the right shows a wind-sea and two swells. Direction here is the direction the waves are travelling \emph{from}. The polar plots have direction from north (rad) on the angular axis and angular frequency (rad Hz) on the radial axis.}
    \label{fig:example_fdspectra}
\end{figure}

Examples of frequency-direction spectra are shown in Figure~\ref{fig:example_fdspectra}, corresponding to: left, wind sea only; middle, wind sea and swell; and right, wind sea with two swells.
Heuristically, if we think about the spectral representation of a process as decomposing that process into a ``sum of sinusoids'', then $S(\omega,\phi)$ can be thought of as describing the contribution to the variance from a wave of a given angular frequency, $\omega$ (measured in rad Hz), travelling from a given direction, $\phi$ (measured in radians).
For example, the left hand panel of Figure~\ref{fig:example_fdspectra} describes a process where most of the variance is generated by sinusoids travelling from a southwards direction ($\pi$ radians) with angular frequencies around 0.8 rad Hz.
In contrast, the right hand example describes a process with major contributions to the variance from sinusoids with three different directions and frequencies.
Notice that the direction is measured clockwise from North in radians and is the direction a wave is travelling \emph{from} and not \emph{towards}.\footnote{Both conventions are used in the literature \citep{Barstow2005}; however, we favour direction from as it means that the relation to the autocovariance is the same as the one used in the statistics literature (see Appendix~\ref{app:fdspec}) and is the same as the convention for wind direction, making comparisons easier.}

Direct characterisation of the wavefield would require measurements of surface displacement over space and time.
Outside of laboratory wave tanks \citep{Forristall2015,Schubert2020}, shallow lakes \citep{Young1996a} or coastal regions \citep{Long1991,Eastoe2013}, this is very difficult to achieve with current technology.
However it is relatively straightforward to measure some characteristics of the wavefield, and to use these measurements to infer properties of the latent spatiotemporal process.
For example, we can use measurements of the motion of a floating buoy to approximate the Lagrangian motion of a particle on the water's surface, recording time series $Z(t), X(t)$ and $Y(t)$, of the vertical, northwards and eastwards displacements of the buoy respectively.

We may also describe the covariance structure of the vector-valued stochastic process $\boldsymbol P(t) = \begin{bmatrix} Z(t) & X(t) & Y(t)\end{bmatrix}^T$
(which is assumed to be stationary and mean-zero) by the spectral density matrix function
\begin{align}
    \boldsymbol{f}(\omega) = 
    \begin{bmatrix}
        f_{zz}(\omega) & f_{zx}(\omega) & f_{zy}(\omega)\\
        f_{xz}(\omega) & f_{xx}(\omega) & f_{yx}(\omega)\\
        f_{yz}(\omega) & f_{xy}(\omega) & f_{yy}(\omega)
    \end{bmatrix}
    = \frac{1}{2\pi}\int_{-\infty}^\infty \boldsymbol c(\tau) e^{-i\omega\tau} \de \tau,
\end{align}
provided certain technical conditions are satisfied (see \citealp{Brockwell2006}, for example), where $\boldsymbol c(\tau)=\EE{\boldsymbol P(\tau)\boldsymbol P(0)^T}$.
Under linear wave theory (see \citealp{Holthuijsen2007}, for example), there is a transfer function
$
    G(\omega,\phi) = \begin{bmatrix} 1 & i\cos\phi & i\sin\phi \end{bmatrix}^T
$ \citep{Isobe1984},
such that
\begin{align}
    \boldsymbol f(\omega) &=
    \int_{0}^{2\pi}
    G(\omega,\phi)G(\omega,\phi)^H
    S(\omega,\phi)
    \de\phi,
    \label{eq:fvsS}
\end{align}
where $A^H$ denotes the conjugate transpose of a matrix $A$. 
This directly relates the frequency-direction spectrum, $S(\omega,\phi)$, to the spectral density matrix function, $\boldsymbol f(\omega)$, which is a feature we shall exploit to perform inference.
From \eqref{eq:fvsS} we can immediately see that, for all $\omega\in\RR$, $\boldsymbol f(\omega)$ is non-negative definite for any non-negative choice of $S(\omega,\phi)$ (and indeed for any choice of $G(\omega,\phi)$, which may be required for other measurement systems, such as a heave-pitch-roll buoy).
Therefore, if we specify a model for $S(\omega,\phi)$ then we can obtain a model for $\boldsymbol f (\omega)$.
However, the relation in \eqref{eq:fvsS} is not invertible in general.

\begin{figure}[h]
    \centering
    \includegraphics[width = \linewidth]{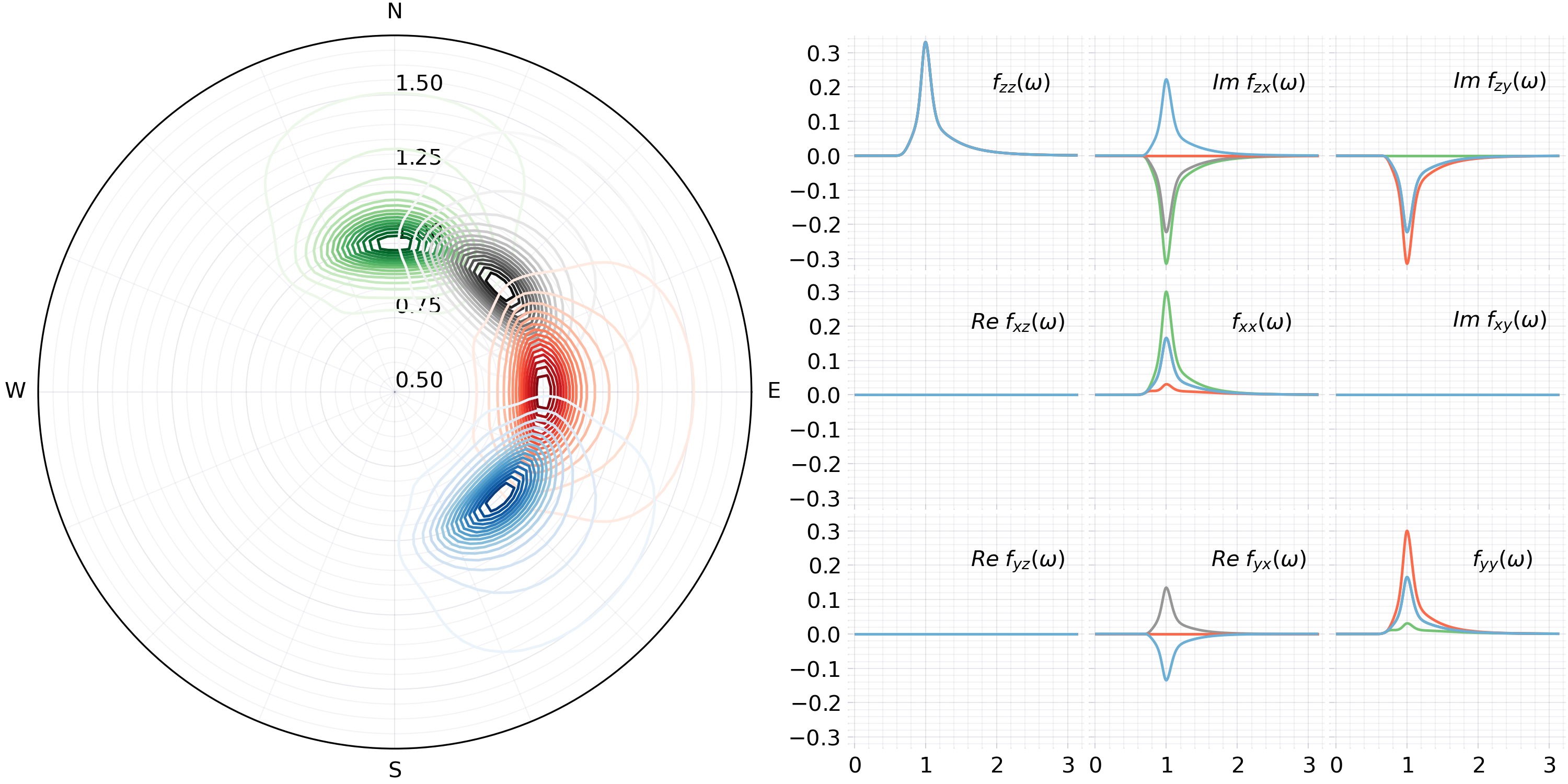}
    \caption{The effect of varying the mean direction of a wind-sea model on both the frequency-direction spectrum ($S(\omega,\phi)$, left) and the corresponding spectral density matrix function ($\boldsymbol f(\omega)$, right). The elements of $\boldsymbol f(\omega)$ are shown as a ``matrix of functions'', plotting the real part in the lower triangle and imaginary part in the upper triangle. Due to the conjugate symmetry of $\boldsymbol f(\omega)$, this representation contains all of the information present.}
    \label{fig:Svsf}
\end{figure}

Figure~\ref{fig:Svsf} shows an example of the relationship between $S(\omega,\phi)$ and $\boldsymbol f(\omega)$ for four different sea states, differing only by mean direction (indicated by the four colours).
The difference in mean direction is obvious in $S(\omega,\phi)$ in the left hand panel, and can still be identified from $\boldsymbol f(\omega)$ in the right hand panel, even though $\boldsymbol f(\omega)$ does not provide a complete description of $S(\omega,\phi)$.

\subsection{Example data}\label{sec:background:example}
For the purpose of illustration, we consider a $z,x,y$ time series recorded using a Datawell Waverider MkIII buoy \citep{Datawell2006} in the southern North Sea, at a sampling rate of 1.28Hz.
This particular five-day period is chosen to provide an illustration of various physical phenomena often present in the ocean.
Within the period, 30-minute sea states (assumed stationary) range from being straightforward to being difficult to model, allowing us to explore the practical applicability of the technique we propose.

\begin{figure}[htp]
    \centering
    \includegraphics[width=\linewidth]{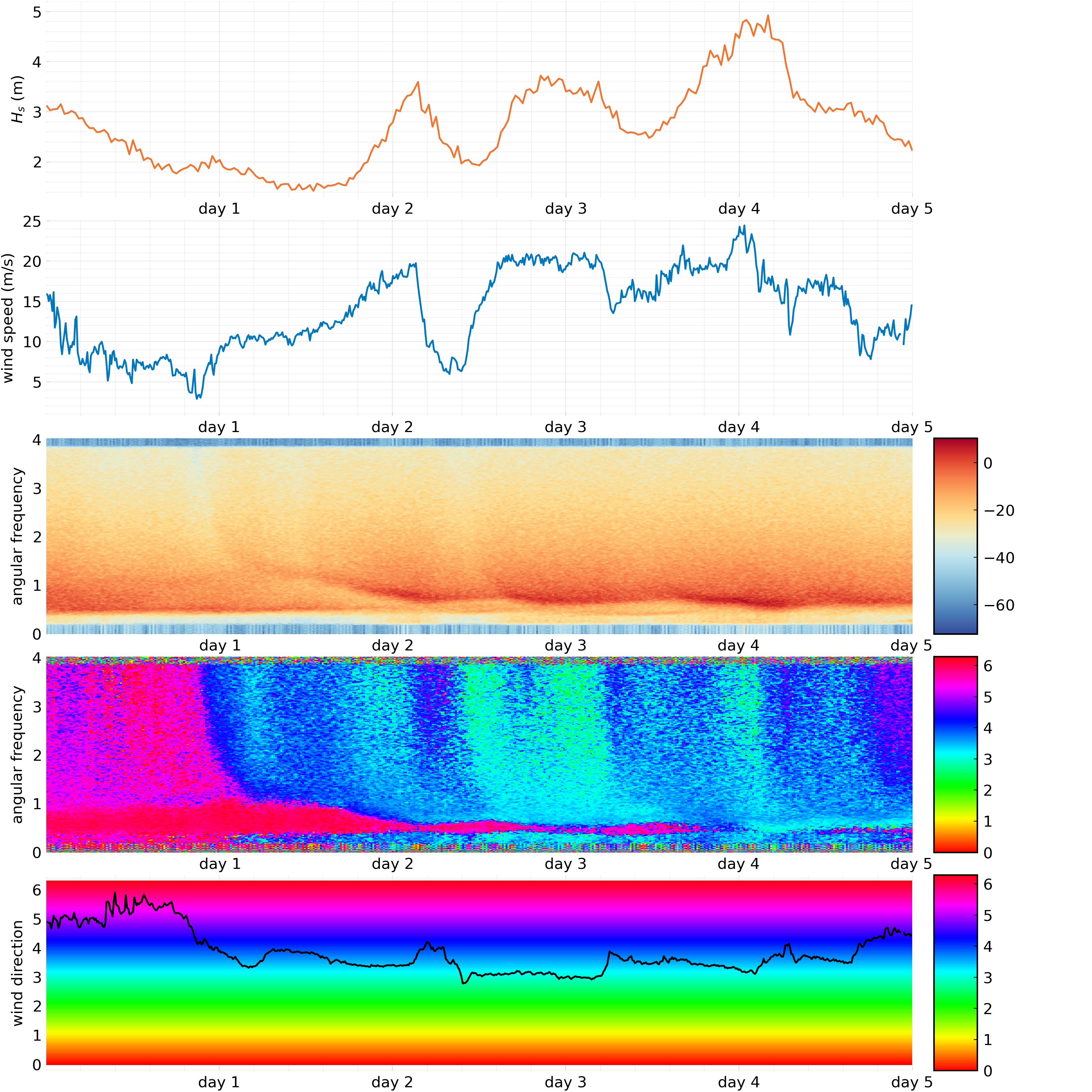}
    \caption{Summary of the storm data, showing significant wave height, wind speed, a spectrogram on the decibel scale, mean wave direction over time and frequency (direction the waves are travelling from, in radians clockwise from North) and the wind direction over time (direction the wind is travelling from, in radians clockwise from North) recorded at a nearby platform. The $x$-axis labels are at the start of the day, e.g. day 1 denotes the start of day 1.}
    \label{fig:spectrogram}
\end{figure}

Figure~\ref{fig:spectrogram} shows a summary of the five-day period in question.
The first panel of Figure~\ref{fig:spectrogram} shows significant wave height, $H_s=4\sqrt{\var{Z(t)}}$, for each of the sea states, quantifying the roughness of the ocean's surface.
The second panel shows wind speed recorded at a nearby platform.
The third panel shows a spectrogram plotted on the decibel scale, computed using multitapering \citep{Thomson1982} with half overlapped 30-minute windows, describing the time-frequency characteristics of the record.
The fourth panel shows the mean direction of the waves at different frequencies over time, as defined by \cite{Kuik1988}, again computed using half overlapped 30-minute windows.
The final panel shows the wind direction over time at a nearby platform.

The record is made up of a variety of component weather systems, which are most easily identified from the mean wave direction (fourth panel).
At the start of the record there are two components present, a high-frequency wind sea and lower-frequency swell.
These components fade out throughout day 0, as can be seen from $H_s$ (first panel). 
A new high-frequency wind sea develops from the start of day 1, with a clear change in mean wave direction (fourth panel).
Throughout day 1, this new component increases in magnitude and transitions to lower frequencies (see third panel), peaking at the start of day 2.
Half way through day 2, the wind drops off and then increases (second panel) and changes direction (fifth panel).
In response, we see another wind-sea component develop with a different direction to the previous component (fourth panel).
Towards the end of day 3, a similar event occurs (though to a lesser degree) and we again see a change in direction.
Meanwhile, the swell persists in a low frequency band throughout (third and fourth panels), though it is small in magnitude and narrow-banded in frequency compared to the wind sea (as can be seen from the spectrogram).

\subsection{Models for wind-sea}\label{sec:background:model}
\begin{figure}[h]
    \centering
    \includegraphics[width = \linewidth]{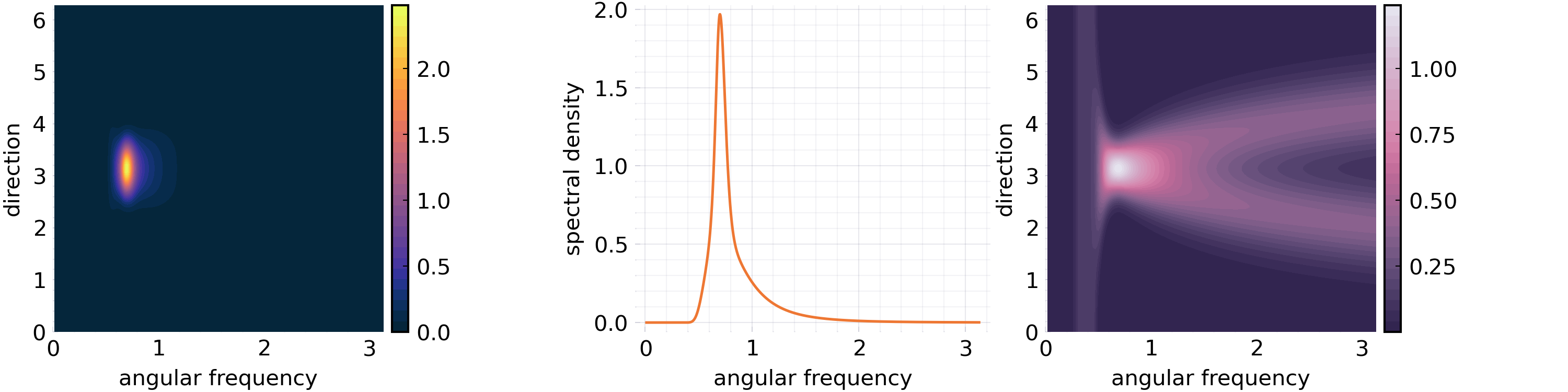}
    \caption{Example of the decomposition of a frequency-direction spectrum, showing the frequency-direction spectrum (left), marginal spectral density function (middle) and spreading function (right).
    Plots are given using Cartesian coordinates as this makes the arms of the spreading function easier to visualise.}
    \label{fig:freqdirdecomp}
\end{figure}

When modelling the frequency-direction spectrum, the spectrum is decomposed as 
\begin{align}
    S(\omega,\phi) = f(\omega)D(\omega,\phi)\label{eq:decomp}
\end{align}
where $f(\omega)$ is the marginal spectral density function of the vertical displacement and $D(\omega,\phi)$ is the so called spreading function.
The marginal spectral density function, $f(\omega)$, can be thought of as describing the contribution to the variance from waves of a given frequency regardless of direction.
Whereas the spreading function, $D(\omega,\phi)$, describes the distribution of wave variance for waves of a given frequency over direction.
The spreading function satisfies
$
    \int_{0}^{2\pi} D(\omega,\phi) \de\phi = 1
$
and $D(\omega, 0) = D(\omega, 2\pi)$ for all $\omega\in\RR$.
Figure \ref{fig:freqdirdecomp} shows an example of the decomposition given in \eqref{eq:decomp} for the model described in the remainder of this section.

For the purpose of this paper, we use the JONSWAP spectral density function first described by \cite{Hasselmann1973}, which we denote $f(\omega;\theta)$, where $\theta$ is the vector of parameters.
The JONSWAP spectral density function is widely used for modelling the univariate vertical surface displacement resulting from wind-sea waves.
Based on physical observations, \cite{Hasselmann1973} developed the JONSWAP spectral density function with an asymmetric peak and a polynomial decay in the high frequency tail.
There is debate about the rate of this tail decay (e.g. \citealp{Hasselmann1973,Toba1973,Phillips1985,Hwang2017}), and so we treat the tail decay index as a free parameter in our analysis.
The form of the JONSWAP spectral density function for ocean surface gravity waves can also be motivated from basic physical considerations. 
Wind waves are generated by the wind blowing over the ocean’s surface, by a combination of three physical processes. Wind field turbulence disturbs the water’s surface, creating high frequency surface water waves. 
Then, wind-wave interaction causes these surface waves to grow in amplitude. 
Thereafter, wave-wave interactions propagate wave energy from higher to lower frequencies. 
This produces a wave spectral density function with a single spectral peak and long high-frequency tail, with peak frequency evolving from higher to lower frequency during an ocean storm of limited duration.

\begin{figure}[h!]
    \centering
    \includegraphics[width = \linewidth]{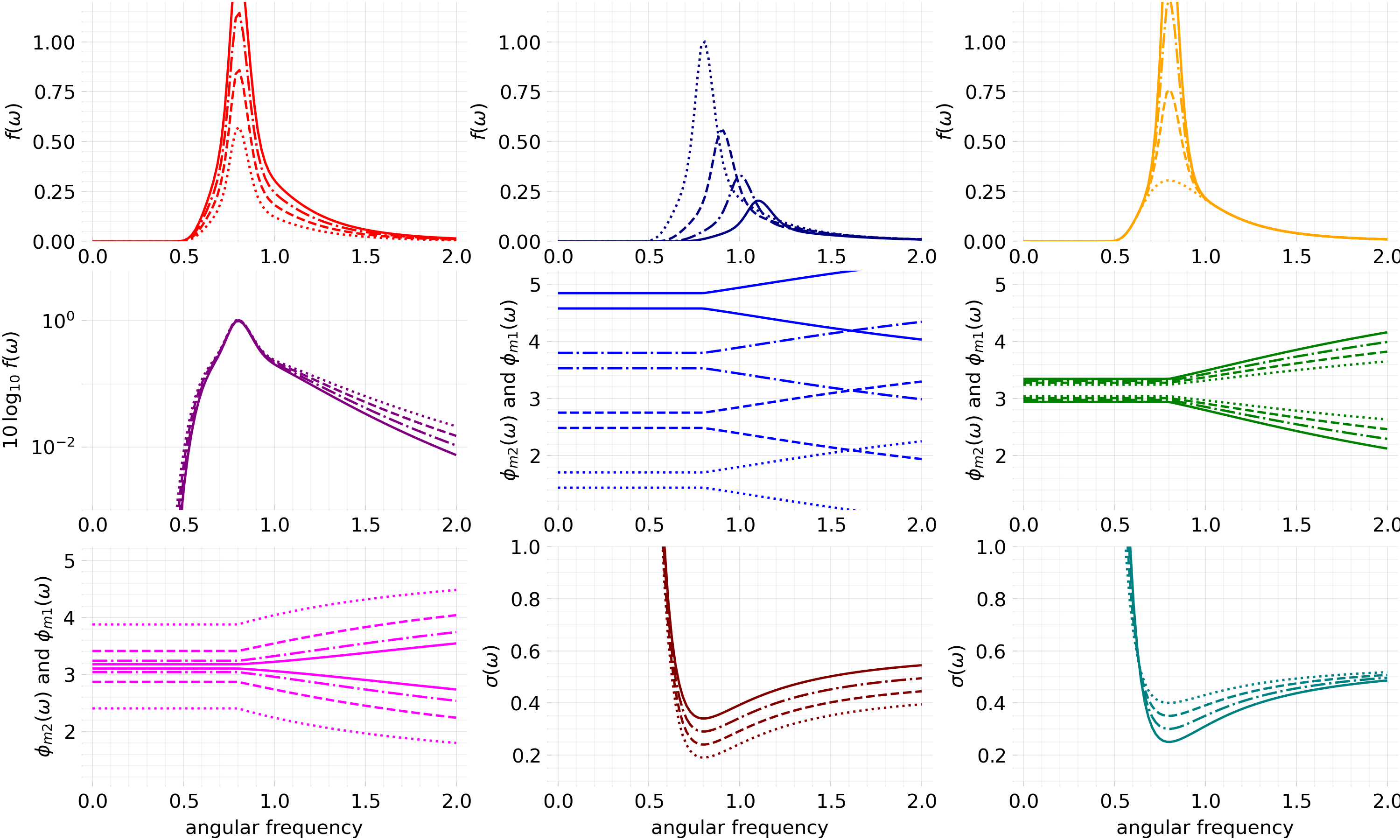}
    \caption{
        The effect of varying different parameters on different components of the model.
        One parameter is varied while the others are held constant.
        The parameters being varied are as follows. 
        Top row: the effect of $\alpha$, $\omega_p$, $\gamma$ on the marginal spectral density function, $f(\omega)$.
        Middle row: the effect of $r$ on $10\log_{10} f(\omega)$; $\phi_m$, $\beta$ on the mean functions, $\phi_{m1}(\omega)$ and $\phi_{m2}(\omega)$.
        Bottom row: the effect of $\nu$ on the mean functions; $\sigma_l$, $\sigma_r$ on the standard deviation function, $\sigma(\omega)$.
        Parameters are held constant with the values $\alpha=0.7$, $\omega_p=0.8$, $\gamma=3.3$, $r=5$, $\phi_m=\pi$, $\beta=4$, $\nu=2.7$, $\sigma_l=0.55$ and $\sigma_r=0.26$.
        Lines become progressively more solid as the value of the parameter increases, and we use values: $\alpha\in\{0.4,0.6,0.8,1.0\}$; $\omega_p\in\{0.8,0.9,1.0,1.1\}$; $\gamma\in\{1,2.5,4,5.5\}$; $r\in\{4,4.5,5,5.5\}$; $\phi_m\in\{\pi/2,5\pi/6,7\pi/6,3\pi/2\}$; $\beta\in\{3,4,5,6\}$; $\nu\in\{1,2,3,4\}$; $\sigma_l\in\{0.45,0.5,0.55,0.6\}$; $\sigma_r\in\{0.15,0.2,0.25,0.3\}$.
    }
    
    \label{fig:vary_parameters}
\end{figure}

Various models have been proposed for the directional spreading of wind-sea waves. 
A large number of experimental studies (e.g. \citealp{Young1995, Ewans1998, Wang2001}) indicate that the spreading function is bimodal with direction, for frequencies exceeding the peak frequency. 
This finding is supported by theoretical arguments involving directional energy transfer through wave-wave interactions \citep{Banner1994,Young1995,Toffoli2010}. 
For this reason, we adopt the bimodal wrapped Gaussian model of \cite{Ewans1998} in this work.
At each frequency, the spreading function over direction is assumed to be a bimodal wrapped Gaussian with means $\phi_{m1}(\omega;\theta)$ and $\phi_{m2}(\omega;\theta)$, but the same standard deviation\footnote{The standard deviation is referred to as angular width by \cite{Ewans1998}.} $\sigma(\omega;\theta)$.
In other words,
\begin{align}
    D(\omega,\phi;\theta) &= \frac{1}{2\sigma(\omega;\theta)\sqrt{2\pi}}\sum\limits_{k=-\infty}^{\infty} \sum\limits_{i=1}^{2} \exp\left\{-\frac{1}{2}\left(\frac{\phi-\phi_{mi}(\omega;\theta)-2\pi k}{\sigma(\omega;\theta)}\right)^2\right\}.\label{eq:bimodalwrappedGaussian}
\end{align}
Table~\ref{tab:parameter_description} gives a description of the parameters of the model, as well as the equations to which they relate, while Figure~\ref{fig:vary_parameters} shows the effect of varying these parameters on the relevant functions.
A complete description of the model is given in Appendix \ref{app:fdspec}.
Note that the inference approach described in this paper is applicable for other models, but the model described here has been chosen for definiteness.

\renewcommand{\arraystretch}{1.2} 
\begin{table}[h!]
    \begin{tabular}{cccl}
        Parameter & Description & Parameter space & \multicolumn{1}{c}{Equation}\\
        \midrule 
        \rowcolor{gray!10} 
        $\alpha$   & scaling parameter & $(0,\infty)$ & \\
        \rowcolor{gray!10} 
        $\omega_p$ & peak frequency & $(0,\infty)$
        & \multirow{-2}{*}{$\;\;\;\;f(\omega;\theta) = 
            \alpha\omega^{-r} \exp\left \{-\tfrac{r}{4}\left(\tfrac{|\omega|}{\omega_p}\right)^{-4}\right \}\gamma^{\delta(\omega;\,\theta)}$} \\
        \rowcolor{gray!10}
        $\gamma$   & peak enhancement factor & $[1,\infty)$ &\\
        \rowcolor{gray!10} 
        $r$        & spectral tail decay index & $(1,\infty)$ & \multirow{-2}{*}{$\;\;\;\;\delta(\omega; \theta)=\exp\left\{-\tfrac{1}{2 (0.07+0.02\cdot\mathbb{1}_{\omega_p>|\omega|})^2}\left (\tfrac{|\omega|}{\omega_p}-1\right )^2\right\}$}\\
        $\phi_m$   & mean direction & $[0,2\pi)$ & \\
        $\beta$    & limiting peak separation & $[0,2\pi)$ & \\
        $\nu$      & peak separation shape & $[0,\infty)$ & \multirow{-3}{*}{$\begin{aligned}\phi_{m1}(\omega;\theta) &= \phi_m + \beta \exp\{-\nu \cdot \min(\omega_p/|\omega|,1)\}/2 \\ \phi_{m2}(\omega;\theta) &= \phi_m -
            \beta \exp\{-\nu \cdot \min(\omega_p/|\omega|,1)\}/2\end{aligned}$}\\
        \rowcolor{gray!10} 
        $\sigma_l$ & limiting angular width & $[0,\infty)$ & \\
        \rowcolor{gray!10} 
        $\sigma_r$ & angular width shape & $[0,\infty)$ & \multirow{-2}{*}{$\;\;\;\;\sigma(\omega;\theta) = \sigma_l - \frac{\sigma_r}{3}\left\{ 4\left(\frac{\omega_p}{|\omega|}\right)^2 - \left(\frac{\omega_p}{|\omega|}\right)^8 \right\}$}
    \end{tabular}
    \caption[Parameter description]{Parameter descriptions and relevant equations. The first four rows describe the parameters for $f(\omega;\theta)$ whilst the others describe the parameters for $D(\omega,\phi;\theta)$ as described in \eqref{eq:bimodalwrappedGaussian}.}
    \label{tab:parameter_description}
\end{table}

\newpage
\section{Modelling process}\label{sec:modelling}
We aim to jointly estimate all the parameters of Table~\ref{tab:parameter_description}, both marginal and spreading, given a sample of three-dimensional displacement.
In this section, we describe the proposed inference technique, and demonstrate in simulations that it yields significant improvements in performance over the existing least squares and moments-matching approaches, described in Appendix~\ref{app:alternatemethods}.
For brevity, we shall refer to such techniques as competitor techniques for the remainder of this paper.
In contrast to competitor techniques, we convert the model for the frequency-direction spectrum to a model for the spectral density matrix function of the data we actually observe, and then fit the model directly to the data. 
This is statistically more appealing as we fully exploit the degrees of freedom in the observational data, rather than performing unnecessary smoothing transformations before model fitting, and is the key reason our method performs better.

\subsection{Model fitting}\label{sec:modelling:fitting}
Due to the quantity of available data, computationally efficient inference techniques are desirable.
For a Gaussian process, full maximum likelihood would require the inversion of a $3n\times 3n$ matrix. 
This is expensive when $n=2304$ as in our case, especially given that we have a different time series every half an hour.
Furthermore, we may wish to only model a certain frequency range (see e.g. Section~\ref{sec:example_analysis:misspecification} for our application study), which is hard to achieve with full maximum likelihood.
Frequency domain psuedo-likelihoods such as the debiased Whittle likelihood \citep{Sykulski2019} provide a nice alternative to full maximum likelihood inference.
Debiased Whittle likelihood inference has been shown to perform well in a variety of applications, including for planetary topography \citep{Guillaumin2019}, ocean drifters \citep{Sykulski2016} and univariate recordings of ocean waves \citep{Grainger2021}.
For these reasons, we use a multivariate extension of the debiased Whittle likelihood due to \cite{Guillaumin2019}.

Let $\boldsymbol P_{t\Delta} = \boldsymbol P(t\Delta)$ for $t \in \ZZ$ be the discrete time process arising from sampling $\{\boldsymbol P(t) \}$ every $\Delta$ seconds. Assume we have a sample of length $n$, the periodogram, $\boldsymbol I_n(\omega)$, is defined as
\begin{align*}
    \boldsymbol I_n(\omega)&=\boldsymbol J_n(\omega) \boldsymbol J_n(\omega)^H \quad \text{where} \quad \boldsymbol J_n(\omega) = \sqrt{\frac{\Delta}{2\pi n}}\sum_{t=0}^{n-1} \boldsymbol{P}_{t\Delta}e^{-it\Delta\omega},
\end{align*}
usually evaluated at the Fourier frequencies $\Omega_n=\{2\pi j/n \mid j\in\{-\lceil n/2\rceil+1,\ldots,\lfloor n/2\rfloor\}\}$ using the Fast Fourier Transform \citep{Cooley1965}.
The multivariate Whittle likelihood \citep{Whittle1953a}, in its discrete form, is given by
\begin{align}
    \ell_W(\theta) &= -\sum_{\omega\in\Omega}\log\left|\boldsymbol f(\omega; \theta)\right| + \tr\{\boldsymbol I_n(\omega) \boldsymbol f(\omega; \theta)^{-1})\},\label{eq:whittle}
\end{align}
where $\Omega \subseteq \Omega_n$ and $ \boldsymbol f(\omega;\theta)$ denotes a parametric spectral density matrix function with parameter vector $\theta$.
The multivariate Whittle likelihood suffers from finite sample bias, especially as the dimension grows, so a debiased version may be used to improve estimates, accounting for finite sampling effects such as aliasing and blurring.
The multivariate debiased Whittle likelihood \citep{Guillaumin2019} is
\begin{align}
    \ell_{D}(\theta) &= -\sum_{\omega\in\Omega}\log\left|\EE{\boldsymbol I_n(\omega); \theta}\right| + \tr\{\boldsymbol I_n(\omega) \EE{\boldsymbol I_n(\omega); \theta}^{-1}\},\label{eq:debiased}
\end{align}
where the expected periodogram can be efficiently computed using the relation
\begin{align*}
    \EE{\boldsymbol{I}(\omega); \theta} &= \frac{\Delta}{2\pi}\sum_{\tau=-n+1}^{n-1} (1-|\tau|/n)\boldsymbol{c}(\tau; \theta) e^{-i\omega\Delta\tau}. 
\end{align*}
In our case, the autocovariance, $\boldsymbol{c}(\tau; \theta)$, is not known analytically, and instead must be approximated numerically from the spectral density matrix function.
Since models are specified for the continuous time process, the most efficient way to approximate the autocovariance is to first approximate the spectral density of the discrete time process, then approximate the autocovariance \citep{Grainger2021}.
The first step requires aliasing the spectral density of the continuous time process by wrapping in contributions from infinitely many frequencies above the Nyquist frequency.
To do this numerically, we have to use a truncated version of this sum.
In practice, the instrument may not respond to waves with frequencies above a certain threshold, or the data may have been filtered in preprocessing.
Therefore, the recorded process may not be aliased to the same extent as the theoretical sampled process.
In our case, we treat the buoys as if no aliasing has occurred (due to the aforementioned preprocessing); however, we note that this technique is able to account for aliasing, should it be felt that aliasing is present.

In both \eqref{eq:whittle} and \eqref{eq:debiased}, summation is over a set $\Omega$. 
Usually $\Omega = \Omega_n$; however, we may wish to remove some frequencies to avoid model misspecification (see Section~\ref{sec:example_analysis:misspecification}) or because at some frequencies in the periodogram the ordinates can be highly correlated for finite samples, which harms Whittle estimation.
We then maximise this likelihood function using numerical methods, detailed further in Appendix~\ref{app:optimisation}.
\newpage
\subsection{Simulation study}\label{sec:modelling:simstudy}
\begin{figure}[h]
    \centering
    \includegraphics[width = \linewidth]{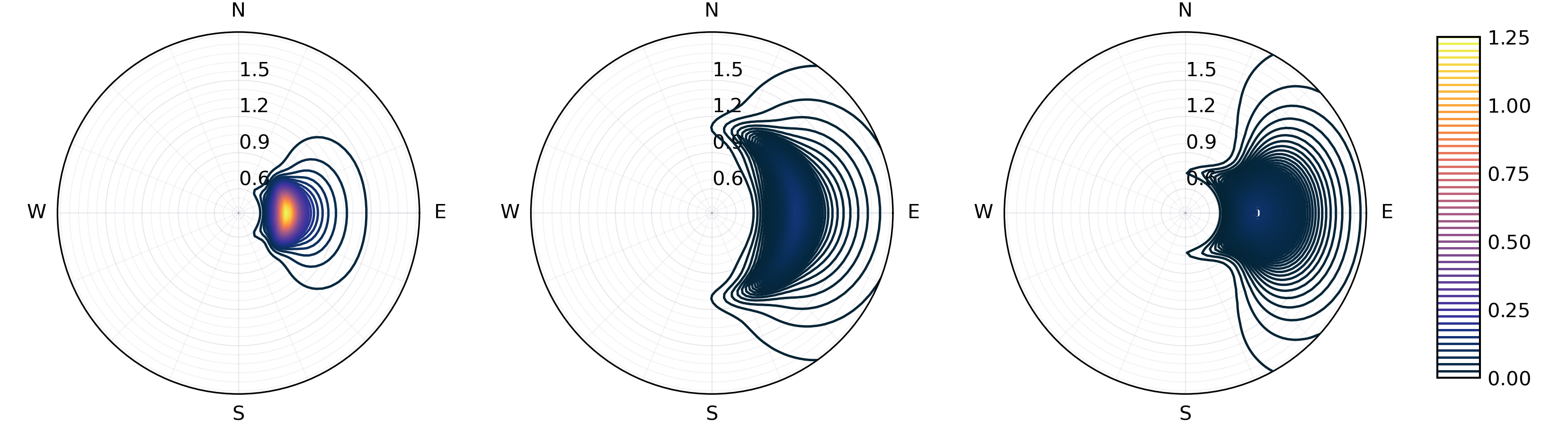}
    \caption{Frequency-direction spectra for Scenario 1 (left), Scenario 2 (middle) and Scenario 3 (right), as defined in in Table~\ref{tab:sim_parameters}.}
    \label{fig:scenario_model}
\end{figure}

We now present a simulation study comparing the debiased Whittle likelihood inference proposed in Section \ref{sec:modelling:fitting} with the least squares and moments-matching techniques described in Appendix~\ref{app:alternatemethods}.
We have chosen three different scenarios that represent possible conditions seen in practice, including cases where certain parameters are on the boundary of the parameter space (as this is likely to cause problems for estimation techniques).
The parameters for each scenario are given in Table~\ref{tab:sim_parameters}, and the corresponding frequency-direction spectra are given in Figure~\ref{fig:scenario_model}.

\begin{table}[ht]
\centering
\begin{tabular}{cccccccccc}
          & $\alpha$ & $\omega_p$ & $\gamma$ & $r$ & $\phi_m$ & $\beta$ & $\nu$ & $\sigma_l$ & $\sigma_r$ \\ \midrule
\rowcolor{gray!10} Scenario 1 & 0.7      & 0.8        & 3.3      & 5   & $\pi/2$  & 4       & 2.7   & 0.55       & 0.26 \\
Scenario 2 & 0.7      & 1.1        & 3.3      & 5   & $\pi/2$  & 4       & 2.7   & 0.55       & 0.00          \\
\rowcolor{gray!10} Scenario 3 & 0.7      & 1.0        & 1.0       & 5   & $\pi/2$  & 4       & 2.7   & 0.55       & 0.26
\end{tabular}
\caption{Table showing the parameters for each scenario in the simulation study.}
\label{tab:sim_parameters}
\end{table}

Scenario 1 is a classic example of a fetch-limited wind sea, with directional shape parameters fixed to the standard values from \cite{Ewans1998}, and $\gamma=3.3$ from \cite{Hasselmann1973}.
Scenario 2 is almost identical, except that $\sigma_r$ is set to $0$, meaning that $\sigma(\omega;\theta)$ is constant over frequency.
Heuristically, this corresponds to a frequency-direction spectrum where the width of each arm in the spreading function is constant over frequency (see Figure~\ref{fig:freqdirdecomp} for the notion of an arm).
This scenario is included because we often see this parameter tending towards the boundary of the parameter space in practice (as in Section~\ref{sec:example_analysis:estimates}) and it is useful to explore the effect of this on other parameter estimates (though we cannot say anything about the impact of model misspecification from this).
Finally, scenario 3 is a Pierson-Moskowitz spectrum for a fully developed sea \citep{Pierson1964}, also using the standard spreading parameters from \cite{Ewans1998}.
This is a special case of the JONSWAP spectrum with $\gamma=1$, and so is of particular interest as it lies on the boundary of the parameter space.

\begin{figure}[htp]
    \centering
    \includegraphics[width = \linewidth]{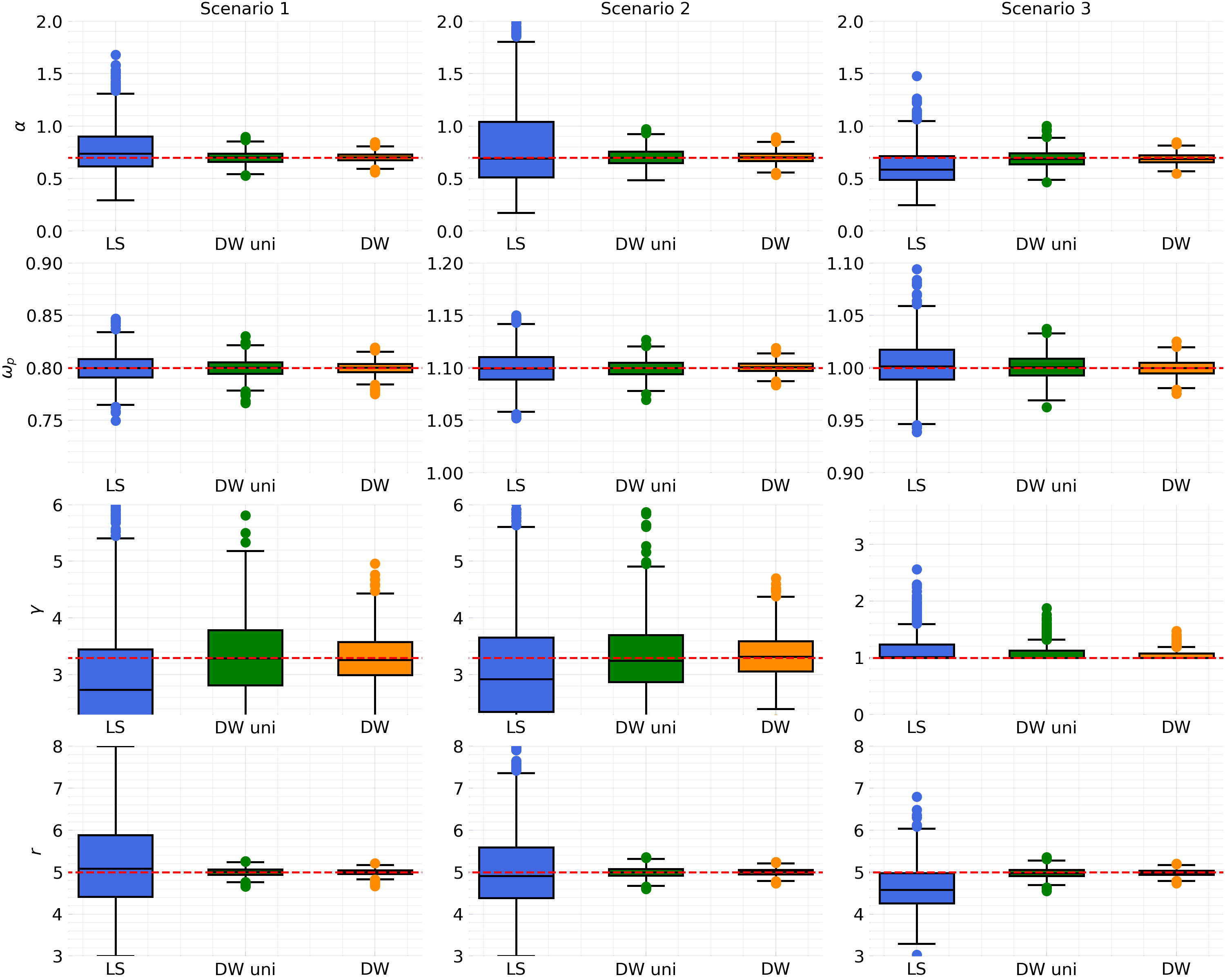}
    \caption{Boxplots of the parameter estimates from the simulation study for parameters of the marginal spectral density function, with the true values indicated by red dashed lines. Marginal parameters estimated using least squares (LS), univariate debiased Whittle (DW uni) and full multivariate debiased Whittle (DW) are shown.}
    \label{fig:scenario_marginal}
\end{figure}

\begin{figure}[htp]
    \centering
    \includegraphics[width = \linewidth]{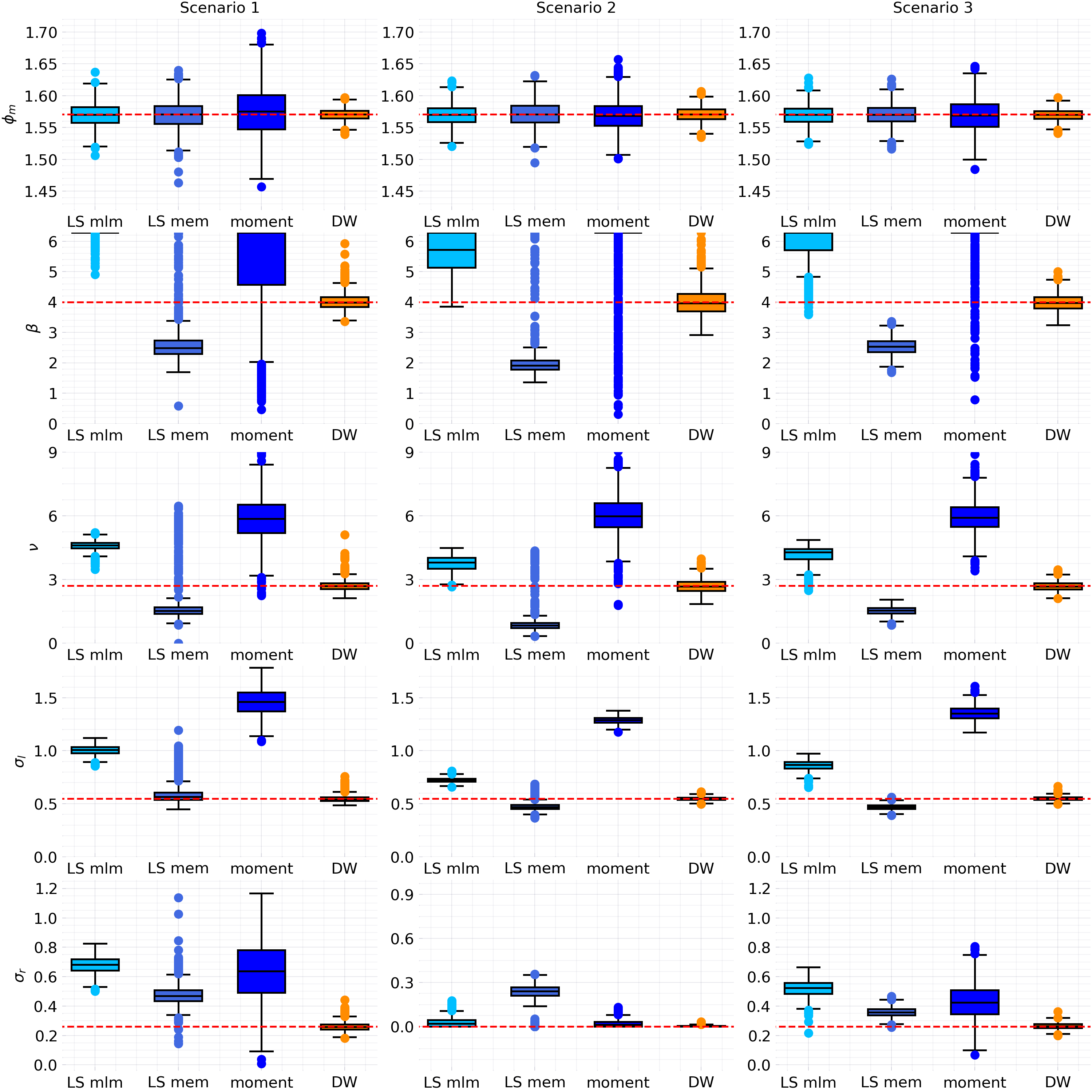}
    \caption{Boxplots of parameter estimates from the simulation study for parameters of the spreading function, with the true values indicated by red dashed lines. 
    Spreading parameters estimated using least squares with MLM (LS mlm), least squares with MEM (LS mem), the moments-matching approach (moment) and multivariate debiased Whittle (DW) are shown.}
    \label{fig:scenario_directional}
\end{figure}

We simulate 1000 time series from each of the scenarios and estimate the model parameters using each of the techniques from Appendix~\ref{app:alternatemethods} alongside the debiased Whittle likelihood inference from Section~\ref{sec:modelling:fitting}.\footnote{In scenario 2, for nine of the replications, the least squares with MEM optimisation did not converge. For this reason, in the results for scenario 2 we include only the 991 replications for which the optimisation of all objective functions converged.}
In particular, we use the least squares technique described in Appendix~\ref{app:alternatemethods:LS} with both MLM and MEM based estimation of frequency-direction spectrum and the moments-matching approach described in Appendix~\ref{app:alternatemethods:moments}.
Whilst there are three different methods from the existing literature in our comparison, they all use the same technique to estimate the parameters of the marginal spectral density function.
As such, Figure~\ref{fig:scenario_marginal} shows the marginal parameters estimated using least squares (the marginal technique for the competitor techniques), univariate debiased Whittle on only the vertical displacement, and multivariate debiased Whittle on all three time series.
Figure~\ref{fig:scenario_directional}, shows the spreading parameters where the least squares technique is now split into the three directional variants: least squares with MLM, least squares with MEM, and the moments-matching approach; and the univariate debiased Whittle is not included (as it cannot be used to estimate the spreading parameters).

From Figure~\ref{fig:scenario_marginal}, we see a clear improvement in the debiased Whittle when comparing to least squares, especially in terms of variance, as has already been reported by \cite{Grainger2021}.
Additionally to the results already seen in \cite{Grainger2021}, there is also a benefit to estimating the parameters of the marginal spectral density function from all three series (as opposed to from the vertical displacement alone).
Traditionally, estimating the marginal parameters has been treated as a separate problem from estimating the spreading parameters, with only the vertical displacements used to estimate the marginal parameters.
However, this clearly throws away useful information about the marginal parameters which is present in the $x$ and $y$ time series.
Furthermore, in Scenario 3, debiased Whittle likelihood recovers $\gamma$ well, despite the true value being on the boundary of the parameter space (though clearly the estimates are not normally distributed).

Similarly, Figure~\ref{fig:scenario_directional} demonstrates a stark difference between competitor techniques and debiased Whittle likelihood inference.
Other than the mean direction $\phi_m$, we see substantial bias in all the other parameter estimates from each of the three existing techniques which is not present in the debiased Whittle likelihood estimates.
We also see that the debiased Whittle estimates exhibit significantly less variance across all parameters and scenarios.
From Scenario 2, we see that debiased Whittle likelihood inference still performs well when $\sigma_r$ is on the boundary of the parameter space (though again estimates are not normally distributed).
Additionally, when estimating $\beta$, we see that least squares with MLM in scenario 1 and moments-matching in scenarios 2 and 3 the majority of the estimates are on the upper boundary of the parameter space, an issue which debiased Whittle likelihood inference does not have.

\newpage

\section{Modelling the example data set}\label{sec:example_analysis}
We now apply debiased Whittle inference for $S(\omega,\phi;\theta)$ (Table \ref{tab:parameter_description}) to the data set introduced in Section~\ref{sec:background:example}.
Both wind sea and swell are present in our example record.
However, we have chosen to model only the wind sea as the purpose of this paper is to introduce a new inference technique, and this is easiest to illustrate and scrutinise with a simple wind-sea only model. The debiased Whittle procedure could naturally be used on a swell-only model (or indeed a joint wind-sea and swell model) should the swell characteristics be of further interest, but this is reserved for future work.

Due to issues with the measurement device and other contaminating processes, certain frequency regions do not reflect the process we are interested in modelling.
Therefore, careful selection of the frequencies included in the objective function must be performed prior to inference.
Selecting these frequencies is difficult, but there are principled ways to choose them.
In particular, the buoy data does not  accurately represent the data which we are interested in modelling at the lowest and highest frequencies \citep{VanEssen2018}.
As such, we select a low- and high-frequency threshold and use only the frequency interval between the thresholds in our analysis, as we shall now detail in Section~\ref{sec:example_analysis:misspecification}.

\subsection{Specification of low- and high-frequency thresholds for inference}\label{sec:example_analysis:misspecification}
Model misspecification presents a significant challenge for the fitting techniques discussed in this paper.
Such misspecification can be generated in a variety of ways.
Firstly, other component weather systems that we do not want to (or cannot) model may be present.
Secondly, there may be noise due to the buoy not following the true motion of a particle on the water's surface.
Finally, the approximations made by linear wave theory that justify the transfer function in \eqref{eq:fvsS} may not be valid.
All of the aforementioned problems affect some frequencies more than others.
Therefore, we shall remove frequencies that are heavily contaminated before fitting models to the data.
Because we are using a frequency domain pseudo-likelihood, this is easy to do, and essentially just involves omitting the appropriate Fourier frequencies from the likelihood (as discussed in Section~\ref{sec:modelling:fitting}).

\begin{figure}[ht]
    \centering
    \includegraphics[width = \linewidth]{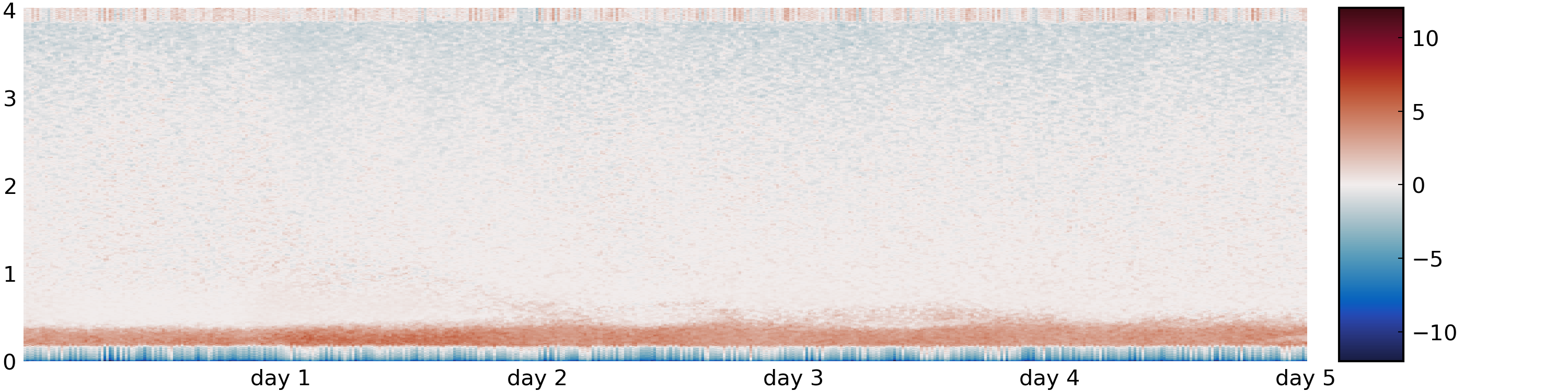}
    \caption{Heatmap of $\hat{R}(\omega)$ for each half hour period in the example data set, computed using multitapering.}
    \label{fig:errorgram}
\end{figure}

However, choosing which frequencies to remove is not trivial.
One useful guide comes from \eqref{eq:fvsS}, which implies that $f_{zz}(\omega) = f_{xx}(\omega) + f_{yy}(\omega)$ under linear wave theory.
Motivated by this, we define the error function $R(\omega) = \log(f_{xx}(\omega)+f_{yy}(\omega)) - \log(f_{zz}(\omega))$.\footnote{Note that this relation is for the deep water case. The finite water depth version is slightly different and given in Appendix~\ref{app:fdspec:finitedepth}. The finite water depth version is used in Figure~\ref{fig:errorgram}, though for simplicity we state the deep water version here. The quantity $R(\omega)$ is related to the check ratio often used in quality control for buoy data \citep{Bushnell2019}.}
An estimate, $\hat{R}(\omega)$, of the error function can be obtained by first estimating the spectral density functions, then plugging them into the above formula for $R(\omega)$. 
Clearly we would expect $\hat R(\omega) \approx 0$ for all $\omega\in[0,\pi/\Delta]$, so deviations from zero may indicate that there is a problem with a certain frequency range.
Figure~\ref{fig:errorgram} shows a plot of $\hat{R}(\omega)$ for each half hour period from our example data set introduced in Section~\ref{sec:background:example} using multitapering.

From Figure~\ref{fig:errorgram}, we see a blue band in the very lowest frequency range with a red band sitting in the frequency range just above this, where the absolute value of the error function is significantly larger than zero.
Therefore, in low frequencies the transfer function mentioned above is not valid, and as a result these frequencies are removed when fitting the model.
Additionally, $\hat R(\omega)$ is slightly negative in the highest frequencies.
In other words, the spectral density of the $\{X(t)\}$ and $\{Y(t)\}$ processes decays more rapidly than that of the $\{Z(t)\}$ process in the high frequency tail.
This is possibly because the accelerometers for measuring the horizontal displacement of the buoy are mounted in a different way to the accelerometer measuring the vertical displacement, though more investigation is needed to ascertain the source of this discrepancy. 
Regardless, it is the general consensus that these instruments are more reliable for the middle of the frequency range than they are at the highest and lowest frequencies, and standard quality control of buoy data include checks for excessive level of low and high frequency spectral density \citep[for example]{Christou2014}.

\begin{figure}[ht]
    \centering
    \includegraphics[width = \linewidth]{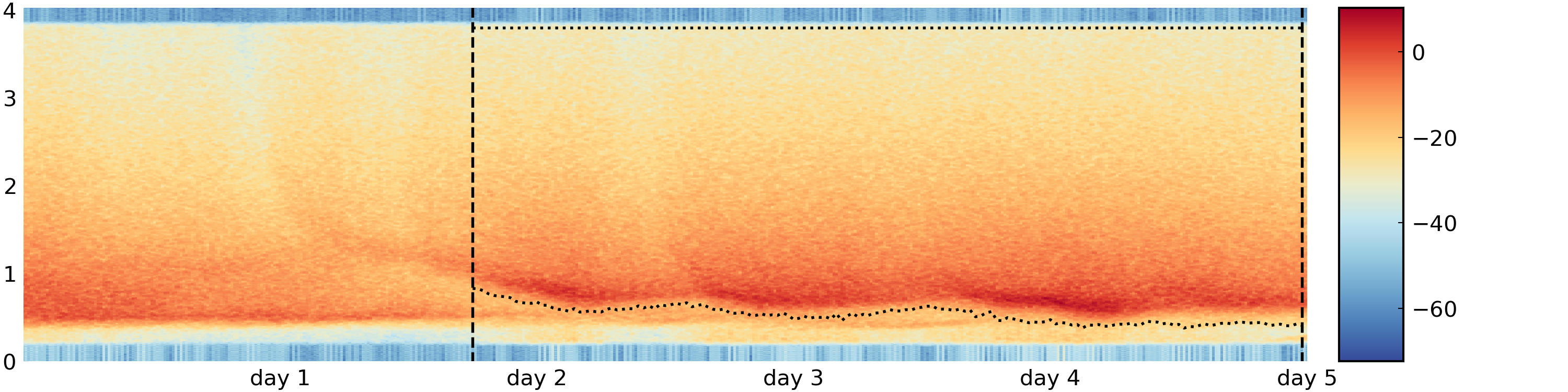}
    \caption{Spectrogram of the example data set on the decibel scale, with the period used in the fitting delimited by solid vertical lines, and the choice of frequency range over the period of interest shown by the dotted lines.}
    \label{fig:threshold}
\end{figure}

Additionally, an old wind sea and a swell are present in the early sea states with the swell persisting, albeit with little energy, for most of the record.
Since models for such conditions are beyond the scope of this paper, we only begin modelling when the new wind sea has become dominant, and remove frequencies in which the swell is large or $\hat{R}$ is sufficiently far from zero.
The specific details are in the code provided on GitHub \url{https://github.com/JakeGrainger/NorthSeaBuoyDisplacementAnalysis.git}.
Figure \ref{fig:threshold} shows this choice of modelling period and low frequency threshold with the modelling period delimited by dashed vertical lines and the threshold shown by dotted lines, indicating that only frequencies between these lines are included.\footnote{Some of the highest frequencies are also removed from the objective function. This is because the response of the buoy falls off rapidly at the highest frequency, which is likely a result of the instrument's inability to respond to the waves and the use of digital filters during post processing.}

\subsection{Parameter estimates}\label{sec:example_analysis:estimates}
Here we estimate model parameters for $S(\omega,\phi;\theta)$ using debiased Whittle inference, for the frequency intervals specified in Section~\ref{sec:example_analysis:misspecification}.
Most of the parameters are initialised from standard values, with the only exceptions being $\omega_p$ and $\phi_m$ which are initialised by picking the frequency corresponding to the maximum of a non-parametric estimate of the marginal spectral density function, and the mean direction corresponding to this frequency respectively.

\begin{figure}[htp]
    \centering
    \includegraphics[width = \linewidth]{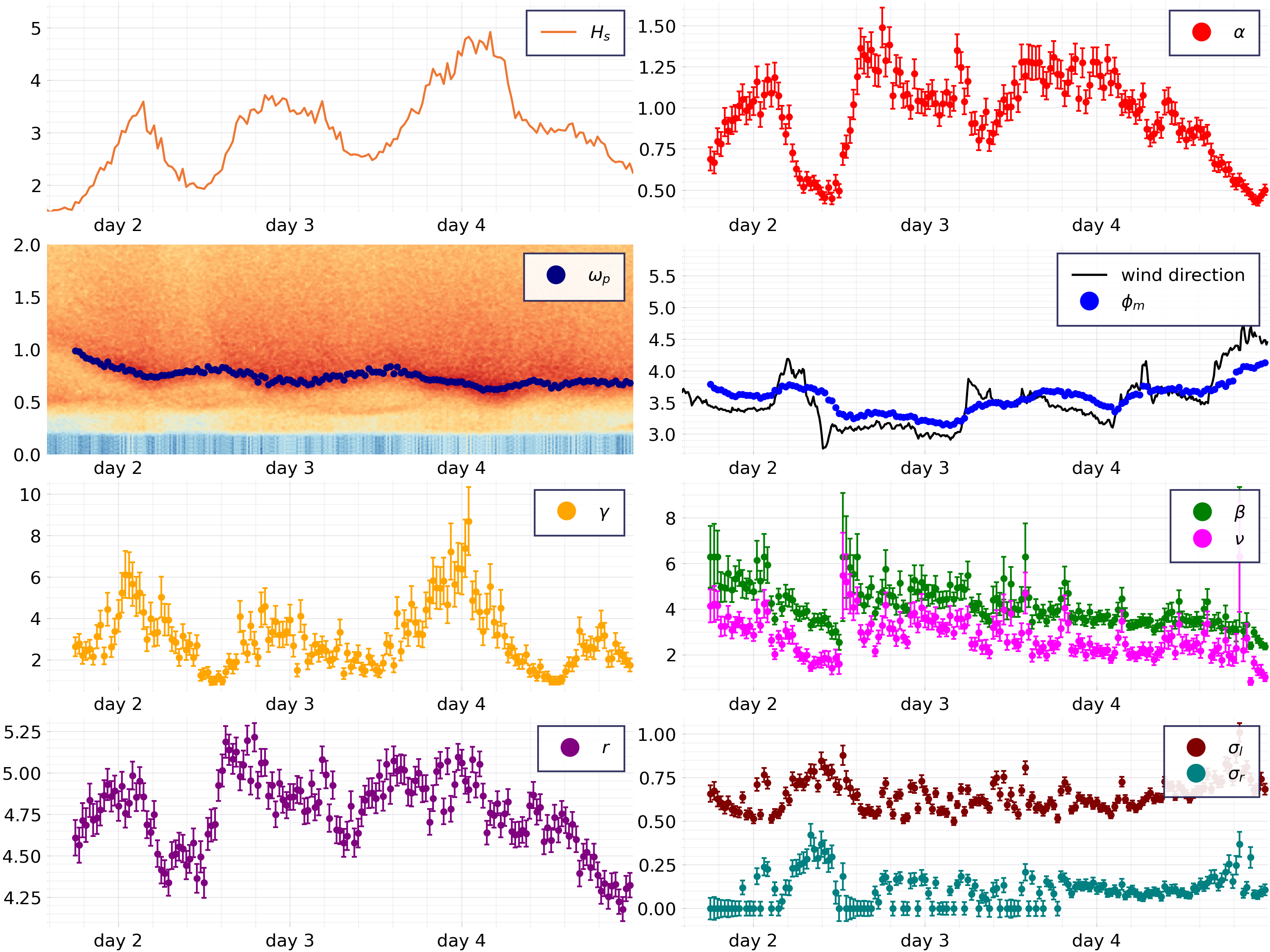}
    \caption{Parameter estimates using debiased Whittle likelihood inference over the period in question with approximate 95\% confidence intervals. The two panels in the second row also include the spectrogram and wind direction for context. In order left to right then down, the panels show: $\hat H_s$; $\hat \alpha$; $\hat\omega_p$ over the spectrogram on the decibel scale; $\hat\phi_m$ and wind direction; $\hat\gamma$; $\hat\beta$ and $\hat\nu$; $\hat r$; and $\hat\sigma_l$ and $\hat\sigma_r$, with approximate 95\% confidence intervals.}
    \label{fig:parest}
\end{figure}

Figure~\ref{fig:parest} shows the parameter estimates, with 95\% approximate confidence intervals, calculated using the expected Hessian matrix and assuming parameters are Gaussian distributed.
The location parameters $\omega_p$ (the peak frequency) and $\phi_m$ (the mean direction) behave as expected, following the spectral mode and reacting to changes in wind-direction respectively.
They also evolve smoothly in time, despite fits being performed independently on non-overlapping sea states.
The shape parameters for the marginal spectral density function ($\gamma$ and $r$), clearly have time varying behaviour. The peak enhancement factor,
$\gamma$, increases as each component wind sea evolves, then decreases as the component wind sea dies out.
Similarly, from the estimates for $r$, the tail decay becomes less steep between components.
It is likely that this is due to model misspecification, as we really have two wind-seas present, but are only modelling one of them.
Furthermore, the shape parameters for the directional spreading ($\beta,\nu,\sigma_l$ and $\sigma_r$) also show anomalous behaviour during these overlaps.
In particular, we see large values of $\beta$ (hitting the upper bound of the parameter space).
Large values of $\beta$ correspond to a wide spreading over direction, which likely occurs because there is another component present with different directional properties.
However, outside these overlap periods we see stability in the parameters estimates, which is encouraging.
Additionally, some of the estimates of $\sigma_r$ drop off to zero, because the low frequency threshold can make $\sigma_r$ unidentifiable as much of the information about $\sigma_r$ resides in frequencies below the peak frequency. 
As a result, we have an identifiability-bias trade off as lowering the threshold frequency introduces more of the noise processes, which tends to result in biasing of $\beta$, but raising the threshold makes $\sigma_r$ unidentifiable.
This is a difficult problem, and is an important area of further research which we discuss more thoroughly in Section~\ref{sec:conclusion}.

In summary, the parameter estimates converge to sensible values in the majority of sea states where a single wind-sea is present. Furthermore, looking at sea states where parameter estimates go to boundaries or unrealistic values helps to extract time periods of interest where the model fails and separate investigation is warranted.

\newpage

\section{Discussion and conclusions}\label{sec:conclusion}

This paper describes estimation of the parameters of frequency-direction spectra for ocean surface gravity waves from three-dimensional buoy displacement time series, using debiased Whittle likelihood inference. 
In simulation studies, debiased Whittle inference is shown to outperform inference using competitor techniques .
Debiased Whittle inference for a sequence of sea states provides a means to characterise the joint evolution of spectral parameters in time, and allows uncertainties in parameter estimates to be quantified in a principled manner.
The observed smooth nature of parameter evolution estimated from North Sea data, and the dependencies evident between parameters, are consistent with physical intuition.

Typically, the wave environment at a location is the product of different physical drivers, including swell and local wind forcing. In the current work, we focus on sea states corresponding to wind-sea conditions only, for clarity of description. 
More generally, debiased Whittle inference for mixed seas consisting of wind-sea and one or more swells is possible; in simulation studies of data for mixed seas (not shown), debiased Whittle inference again performs well.
In simulation studies on samples of 30-minute records corresponding to wind-sea conditions, the debiasing procedure makes a small but marginal improvement over standard multivariate Whittle estimation. However, when fitting the joint wind-sea and swell model to mixed sea states, estimates using standard Whittle inference exhibit substantially greater bias than those from debiased Whittle inference.

In-situ measurement of the ocean environment is invariably problematic. 
In the present study, buoy displacement time series are contaminated by additional low-frequency processes, leading to spurious low-frequency spectral features not represented in the assumed parametric spectral form to be estimated. 
At very high frequencies, buoy displacement time series are further subject to on-board low-pass filtering effects not represented in the assumed spectral form.
We adjust the inference procedure for these sources of model misspecification by only considering a central band of frequencies in the likelihood, set using low-frequency and high-frequency thresholds.
In general, the low-frequency threshold in particular should be chosen carefully, to achieve a good balance between model misspecification (when the threshold is too low) and identifiability (when the threshold is set so high that aspects of the spectral form cannot be resolved). 
We have explored extending the spectral form to accommodate an additional low-frequency noise feature, but found that achieving this reliably required a large number of extra parameters, and resulted in greater loss of efficiency in estimating the wind-sea (and swell) components of interest compared to frequency thresholding.

Spectral estimates in the current work are based on data for the ocean's surface displacement only. In general, it would be advantageous to incorporate the effects of covariates such as the evolving wind field on the spectral form, particularly for characterisation of mixed seas. For example, the direction associated with a wind-sea component at a location is dependent on local wind speed and direction, whereas the characteristics of a swell component do not vary substantially with the local wind field. These covariate dependencies are often exploited by physical oceanographers to partition the frequency-direction domain into sub-domains corresponding to wind-sea and swell components \citep[for example]{Hanson2001}.

The spectral characteristics of ocean waves evolve smoothly in time. 
In this paper, as is common practice, we accommodate temporal non-stationarity by partitioning time series into consecutive 30-minute sea states which are considered stationary for purposes of spectral inference.
Improved bias-variance properties of parameter estimates from debiased Whittle inference suggest that spectral estimation using sea states of shorter duration is feasible for more-rapidly evolving ocean environments; initial simulation studies (not shown) support this finding. 
More generally, simultaneous spectral estimation for a sequence of consecutive sea states exploiting smooth time-varying basis representations for spectral parameters (e.g. using splines), or adaptive estimation of evolving spectral forms (e.g. using dynamic linear models) are obvious research avenues.

The methodology for spectral inference described in this paper is generally applicable, provided that an appropriate model for the spectral density matrix function can be obtained by applying a suitable transfer function to the model for the frequency-direction spectrum. 
Thus, in addition to three-dimensional buoy displacement time series, debiased Whittle inference is applicable to  heave-pitch-roll buoy data, for example.
A collection of useful transfer functions for commonly used oceanographic devices is given by \cite{Benoit1997}.
The methodology can be modified for similar applications involving observations of a process viewed as a linear time-invariant filter of some latent process of interest.
Practical issues encountered in the current work, relating to time series aliasing, unusual sources of measurement noise and complex likelihood functions are common across many applications (e.g. involving accelerometers and GPS tracking).
Hence, we hope that the methodology presented and the ideas it incorporates will prove useful to the practicing oceanographer, ocean engineer and applied statistician.

\subsection*{Acknowledgements}
J. P. Grainger’s research was funded by the UK Engineering and Physical Sciences Research Council (Grant EP/L015692/1) and JBA Trust. 
The work of A. M. Sykulski was funded by the UK Engineering and Physical Sciences Research Council (Grant EP/R01860X/1). 
We thank Allan Rod Zeeberg and TotalEnergies for providing access to the North Sea dataset, for useful discussions and for allowing an anonymised portion of the data to be made available in the name of reproducible research.

\textbf{The environmental data used has been provided by TotalEnergies E\&P Danmark A/S, and the data is the sole property of TotalEnergies E\&P Danmark A/S, and it may not be used or reproduced without the written consent of TotalEnergies E\&P Danmark A/S. TotalEnergies E\&P Danmark A/S has not assisted with or had any influence on the use of the data or the subject or content of this paper.}
\newpage

\appendix
\section{The frequency-direction spectrum}\label{app:fdspec}
\subsection{Definition and relation to autocovariance}\label{app:fdspec:def}
We are interested in modelling the spatiotemporal process $\{\eta(x,y,t)\}_{(x,y,t)\in\RR^3}$, which constitutes the surface displacement of the water over time and space.
Let $\tau$ be the temporal lag, and $\boldsymbol{l} = [l_x,l_y]$ be a vector of spatial lags.
Then, assuming stationarity, $c_\eta(\boldsymbol l, \tau) = \EE{\eta(l_x,l_y,\tau)\eta(0,0,0)}$ is defined to be the autocovariance of the spatiotemporal process.
Denote the angular frequency by $\omega$ and wave-vector by $\boldsymbol\kappa=[k_x,k_y]$.
The spectral density function of the spatiotemporal process is then
\begin{align*}
    f_\eta(\boldsymbol\kappa,\omega) &= \frac{1}{\left(2\pi\right)^3} \int_{\RR}\int _ {\RR^2} c_\eta(\boldsymbol\kappa,\tau) e^{-i(\omega\tau+\boldsymbol{\kappa}\cdot\boldsymbol{l})} \de \boldsymbol l \de\tau 
\end{align*}
Note that this is different to the definition given in some oceanography texts \citep[for example]{Barstow2005}, this is because the angle $\phi=\arg(k_x+ik_y)$ is defined as the direction the wave is coming from, and so we need $\omega\tau+\boldsymbol\kappa\cdot\boldsymbol l$ as opposed to $\omega\tau-\boldsymbol\kappa\cdot\boldsymbol l$ in the exponential function (which is present in definitions when the author wants direction to be the direction the waves are propagating towards).
Of course, this is only a convention, but it is worth noting the difference when applying these techniques.

Now, whilst in general a process such as $\{\eta(x,y,t)\}_{(x,y,t)\in\RR^3}$ requires a spatial spectral density function expressed in terms of a frequency and two wavenumbers (a wave-vector), under a simplification of the governing equations for wave dynamics known as linear wave theory, the absolute value of the wave-vector is specified uniquely by the absolute value of the frequency, using a dispersion relation.
For this reason, the covariance structure of the process is usually specified by a frequency-direction spectrum, which we denote $S(\omega,\phi)$, with the relation
\begin{align*}
    f_\eta([k\cos(\phi), k\sin(\phi)], \omega) &= \begin{cases}
        S(\omega,\phi) & \text{if }\omega^2 = kg\tanh(kh),\\
        0 & \text{otherwise,}
    \end{cases}
\end{align*}
where $h$ denotes the water depth when the water is still (which is assumed constant).
More details can be found in \cite{Barstow2005} and references therein, for example.

\subsection{Models for a wind-sea frequency-direction spectrum}\label{app:fdspec:models}
One of the most widely used spectral density functions for modelling the univariate vertical surface displacement resulting from wind-sea waves is the JONSWAP \citep{Hasselmann1973} spectral density function, given by
\begin{align*}
f(\omega;\theta)&=
    \begin{cases}
        \alpha\omega^{-r}\exp\left \{-\frac{r}{4}\left(\frac{|\omega|}{\omega_p}\right)^{-4}\right \}\gamma^{\delta(|\omega|; \theta)} & \text{for }|\omega|>0,\\
        0 & \text{for }\omega=0.
    \end{cases}
\end{align*}
where
\begin{align*}
	\delta(\omega; \theta)=\exp\left\{-\frac{1}{2\sigma(\omega; \theta)^2}\left (\frac{\omega}{\omega_p}-1\right )^2\right\},& \quad&
    \sigma(\omega; \theta)=\begin{cases}
    	0.07\quad\text{for }\omega\leq\omega_p\,,\\
    	0.09\quad\text{for }\omega>\omega_p\,,
    \end{cases}	
\end{align*}
with parameters $\alpha, \omega_p , \gamma, r$.

For the spreading function, we use the bimodal wrapped Gaussian model proposed by \cite{Ewans1998}.
The bimodal wrapped Gaussian spreading function is defined as
\begin{align}
    D(\omega,\phi;\theta) &= \frac{1}{2\sigma(\omega;\theta)\sqrt{2\pi}}\sum\limits_{k=-\infty}^{\infty} \exp\left\{-\frac{1}{2}\left(\frac{\phi-\phi_{m1}(\omega;\theta)-2\pi k}{\sigma(\omega;\theta)}\right)^2\right\}+\exp\left\{-\frac{1}{2}\left(\frac{\phi-\phi_{m2}(\omega;\theta)-2\pi k}{\sigma(\omega;\theta)}\right)^2\right\}\nonumber
\end{align}
for $\omega\in\RR$ and $\phi\in[0,2\pi]$;
where $\phi_{m1}(\omega;\theta)$ and $\phi_{m2}(\omega;\theta)$ are the mean direction functions and $\sigma(\omega;\theta)$ is the angular width function.
These functions are themselves parameterised. 
\cite{Ewans1998} gives a parameterisation with fixed values based on observed data, with a single location parameter to determine the mean direction. 
We shall use a less restrictive description by adding parameters for the shape and scale of the spreading (a similar parametrisation was used by \cite{vanZutphen2008}, but we use slightly fewer parameters as some of the parameters in \cite{vanZutphen2008} have little effect on the frequency-direction spectrum). In particular, we write
\begin{align*}
    \phi_{m1}(\omega;\theta) &= \phi_m + \phi_s(\omega;\theta)/2, \\
    \phi_{m2}(\omega;\theta) &= \phi_m - \phi_s(\omega;\theta)/2, \\
    \phi_s(\omega;\theta) &= 
        \begin{cases}
            \beta \exp(-\nu \omega_p/|\omega|) & \text{for } |\omega| > \omega_p, \\
            \beta \exp(-\nu) & \text{otherwise,}
        \end{cases}\\
    \sigma(\omega;\theta) &= \sigma_l - \frac{\sigma_r}{3}\left( 4\left(\frac{\omega_p}{|\omega|}\right)^2 - \left(\frac{\omega_p}{|\omega|}\right)^8 \right).
\end{align*}
This adds an additional 5 parameters, namely $\phi_m,\beta,\nu,\sigma_l,\sigma_r$.

\subsection{Corresponding models for particle displacement}
From \eqref{eq:fvsS} we can see that the corresponding model for the spectral density matrix function of the displacement of a particle in a wind-sea is

\begin{align}
    \boldsymbol f(\omega;\theta) &= f(\omega;\theta)
    \begin{bmatrix}
        1 & -w_{xz}(\omega;\theta) & -w_{yz}(\omega;\theta)\\
        w_{xz}(\omega;\theta) & w_{xx}(\omega;\theta) & w_{xy}(\omega;\theta)\\
        w_{yz}(\omega;\theta) & w_{xy}(\omega;\theta) & w_{yy}(\omega;\theta)
    \end{bmatrix}
\end{align}
where
\begin{align*}
    w_{xx}(\omega;\theta) &= \frac{1}{2}\left(1+\cos(2\phi_m)\cos(\phi_s(\omega;\theta))e^{-2\sigma^2(\omega;\theta)}\right), \quad &w_{xz}(\omega;\theta) = i\cos(\phi_m)\cos(\phi_s(\omega;\theta)/2)e^{-\sigma^2(\omega;\theta)/2},
    \\
    w_{yy}(\omega;\theta) &= \frac{1}{2}\left(1-\cos(2\phi_m)\cos(\phi_s(\omega;\theta))e^{-2\sigma^2(\omega;\theta)}\right), \quad &w_{yz}(\omega;\theta) = i\sin(\phi_m)\cos(\phi_s(\omega;\theta)/2)e^{-\sigma^2(\omega;\theta)/2}, \\
    w_{xy}(\omega;\theta) &= \frac{1}{2}\sin(2\phi_{m})\cos(\phi_{s}(\omega;\theta))e^{-2\sigma^2(\omega;\theta)}. &
\end{align*}
\subsection{Finite water depth correction}\label{app:fdspec:finitedepth}
The relation given in \eqref{eq:fvsS} is for waves in deep water.
For finite water depths a slightly different relation is required.
In particular, we have 
$$G(\omega,\phi) = \begin{bmatrix} 1 & i\cos\phi/\tanh(kh) & i\sin\phi/\tanh(kh) \end{bmatrix}^T$$
where $h$ is the water depth (in metres)\footnote{In our case, $h\approx40$.} and $\omega^2 = k \tanh(kh)$.
Consequently we have

$$f_{zz}(\omega) = (f_{xx}(\omega)+f_{yy}(\omega)) \tanh(kh)^2$$

meaning that the correct definition for $R(\omega)$ is

$$R(\omega) = \log(f_{xx}(\omega)+f_{yy}(\omega))+2\log(\tanh(kh))-\log(f_{zz}(\omega)).$$

It is this definition for $R(\omega)$ we use in Figure~\ref{fig:errorgram}.

\section{Alternate inference techniques}\label{app:alternatemethods}
In this appendix, we describe competitor techniques for estimating the parameters of model frequency-direction spectrum from observed buoy data.
Typically, a two stage approach is taken. 
Firstly, the parameters of the spectral density function of the vertical displacement are estimated using least squares\footnote{which has been shown to perform poorly for many parameters of interest \citep{Ewans2018,Grainger2021}.} and then the parameters of the spreading function are estimated separately. 
The latter is usually done in one of two ways: a moments-matching approach \citep[for example]{Ewans1998}; or by producing a non-parametric estimate of the spreading function, then fitting using least squares.
Both of these techniques start from writing the spreading function as a Fourier series (which is possible from the periodicity of the spreading function):
\begin{align*}
        D(\omega,\phi) &= \frac{1}{\pi}\left(\frac{1}{2} + \sum_{n=1}^\infty a_n(\omega)\cos(n\phi)+b_n(\omega)\sin(n\phi)\right)\;.
\end{align*}
From \eqref{eq:fvsS} we can see that
\begin{align*}
    a_1(\omega) &= \frac{\Im(f_{xz}(\omega))}{f_{zz}(\omega)}, &
    b_1(\omega) &= \frac{\Im(f_{yz}(\omega))}{f_{zz}(\omega)}, &
    a_2(\omega) &= \frac{f_{xx}(\omega)-f_{yy}(\omega)}{f_{zz}(\omega)}, &
    b_2(\omega) &= \frac{2f_{xy}(\omega)}{f_{zz}(\omega)}.
\end{align*}
The remaining coefficients cannot be obtained from the cross-spectra in general.
The approach to solving this problem has typically been to guess at the remaining Fourier frequencies either based on the Fourier coefficients of a model or by making some other assumptions about the behaviour of the spreading function.
Of course, this assumes we know the cross-spectra, but we must estimate them.
This distinction is not trivial.

\subsection{Least squares fitting to estimates of the spreading function}\label{app:alternatemethods:LS}
A commonly used technique involves fitting the model spreading function to a non-parametric estimate of the spreading function using least squares. In other words, given $\hat D(\omega,\phi)$, an estimate of the spreading function\footnote{using techniques such as MLM \citep{Isobe1984} and MEM \citep{Lygre1986}.  See \cite{Benoit1997} for a summary of these.}, the parameters are obtained by solving
\begin{equation*}
    \argmin\limits_\theta \sum_{\omega\in\Omega}\sum_{\phi\in\Phi} \left(D(\omega,\phi;\theta)-\hat{D}(\omega,\phi)\right)^2.
\end{equation*}
The problem with this is that such a technique assumes that the estimator used for the spreading function is unbiased, normally distributed, homoscedastic and that at different pairs of frequency and direction estimates are uncorrelated; however, none of these are satisfied in practice. 
In particular, correlation across frequency is high for both MLM and MEM methods, and bias is substantial.
As a result, estimation of anything other than location parameters using this technique performs poorly. 

\subsection{Moments-matching approach}\label{app:alternatemethods:moments}
Early approaches to fitting parametric spreading functions to data from buoys, such as \cite{Mitsuyasu1975}, tended to match the Fourier coefficients estimated from the buoy to the theoretical Fourier coefficients from the model (under the relevant transfer function).
These approaches actually work by first estimating the parameters of the spreading function at each frequency (which are different from the model parameters), and then essentially doing regression analysis to work out the parameters of the model for the behaviour of the spreading function over frequency. 
In our case, following \cite{Ewans1998}, at each frequency estimate $\theta_\omega = [\phi_{m1}(\omega),\phi_{m2}(\omega),\sigma(\omega)]$ using
\begin{equation*}
    \hat\theta_\omega = \argmin\limits_{\theta_\omega} \LSmoments{c}{1} + \LSmoments{c}{2}
\end{equation*}
where $c_j(\omega;\theta)=a_j(\omega;\theta)+i b_j(\omega;\theta)$ and $\hat{c}_j$ is an estimate for $c_j(\omega;\theta)$ obtained by plugging in estimates for the relevant cross-spectra (typically estimated using some variation on Welch's Method), for $j=1,2$.

The parameters of interest $(\beta,\nu,\sigma_l,\sigma_r)$ are then estimated by
\begin{equation*}
    \hat\theta=\argmin\limits_\theta \sum_{\omega\in\Omega} \LSregress{\phi_{m1}} + \LSregress{\phi_{m2}} + \LSregress{\sigma}.
\end{equation*}
Such a technique is usually not applied to a single sea state, but instead is applied to multiple sea states with the view to fixing the parameters of the spreading function (bar the mean direction).
You can apply this to a single sea state but, as we show in Section~\ref{sec:modelling:simstudy}, this performs poorly.
However, it should be remembered that this technique can still be useful for getting a general idea of the shape different aspects of the spreading function can take, but it is not useful for estimating the parameters of a single sea state.

\section{Optimisation and gradient calculation}\label{app:optimisation}
Parameters are estimated jointly by optimising the debiased Whittle likelihood using the interior point Newton method as implemented in Optim.jl \citep{Optim.jl-2018}.
We use Fisher scoring as the expected Hessian of the debiased Whittle likelihood can be computed analytically from the first derivatives of the autocovariance (whilst the Hessian would require the second derivatives as well).
This results in very fast optimisation compared to other approaches.
For further details, see the Julia package WhittleLikelihoodInference.jl (\url{https://github.com/JakeGrainger/WhittleLikelihoodInference.jl.git}).



\bibliography{bib/extra,bib/library}

\end{document}